**No evidence of phosphine in the atmosphere of Venus by independent analyses**

Villanueva G.L.[1], Cordiner, M.[1,2], Irwin P.G.J.[3], de Pater I.[4], Butler B.[5], Gurwell M.[6], Milam S.N.[1], Nixon C. A.[1], Luszcz-Cook S. H.[7,8], Wilson C.F.[3], Kofman V.[1,9], Liuzzi G.[1,9], Faggi S.[1,9], Fauchez T.J.[1,10], Lippi M.[1,9], Cosentino R.[1,11], Thelen A. E.[1,10], Moullet A.[12], Hartogh P.[13], Molter E.M.[4], Charnley S.[1], Arney G.N.[1], Mandell A.M.[1], Biver N.[14], Vandaele A.C.[15], de Kleer K. R.[16], Kopparapu R.[1]

1. NASA Goddard Space Flight Center, Solar System Exploration Division, Greenbelt MD 20771, USA
2. Catholic University of America, Department of Physics, Washington DC, USA
3. University of Oxford, Dept. of Physics, Oxford, UK
4. University of Berkeley, Department of Astronomy, Berkeley CA, USA
5. National Radio Astronomy Observatory (NRAO), Socorro NM, USA
6. Center for Astrophysics, Harvard & Smithsonian, Cambridge, MA, USA
7. Columbia University, New York NY, USA
8. American Museum of Natural History, New York NY, USA
9. American University, Physics Department, Washington DC, USA
10. Universities Space Research Association, Columbia MD, USA
11. University of Maryland, College Park MD, USA
12. SOFIA Science Center, Moffett Field CA, USA
12. Max-Planck-Institut fur Sonnensystemforschung, Gottingen, Germany
13. Observatoire de Paris, LEISA/CNRS, Meudon, France
14. Royal Belgian Institute for Space Aeronomy, BIRA-IASB, Brussels, Belgium
15. California Institute of Technology (Caltech), Pasadena CA, USA

The detection of phosphine ($PH_3$) in the atmosphere of Venus has been recently reported based on millimeter-wave radio observations[1] (G2020 hereafter), and its re-analyses[2,3] (G2021a/b thereafter). In this Matters Arising we perform an independent reanalysis, identifying several issues in the interpretation of the spectroscopic data. As a result, we determine sensitive upper-limits for $PH_3$ in Venus' atmosphere (>75 km, above the cloud decks) that are discrepant with the findings in G2020 and G2021a/b.

The measurements target the fundamental first rotational transition of $PH_3$ (J=1-0) at 266.944513 GHz, which was observed with the James Clerk Maxwell Telescope (JCMT) in June 2017 and with the Atacama Large Millimeter/submillimeter Array (ALMA) in March 2019. This line's center is near the $SO_2$ (J=$30_{9,21}$-$31_{8,24}$) transition at 266.943329 GHz (only 1.3 km/s away from the $PH_3$ line) which represents a potential source of contamination. The JCMT and ALMA data, as presented in G2020, are at spectral resolutions comparable to the frequency separation of the two lines. Moreover, the spectral features identified are several km/s in width, and therefore do not permit distinct spectroscopic separation of the candidate spectral lines of $PH_3$ and $SO_2$. We present the radiative transfer modelling we have performed and then discuss the ALMA and JCMT analyses in turn.

**ALMA reanalysis**: The analysis of interferometric data is relatively complex, in particular for bright and extended sources such as Venus (15.2 arcsecs angular diameter for the ALMA data). The completeness of the different baselines (short and long) determines the ability to accurately measure the total planetary flux density[9], while the bandpass calibration is a crucial factor in the ultimate quality of the resulting spectra[10,11]. The details of the calibration and imaging used during data reduction can have a dramatic impact on the quality and validity of the resulting ALMA interferometric data. The extracted spectra in Extended Figure 4" and the interferometric map in Extended Figure 3 of G2020 show large quasi-periodic fluctuations in the spectrum, which in G2020 is fitted with high-order polynomials. Particularly challenging is the fact that these fluctuations have a pattern/width comparable to their defined $PH_3$ line core region. As we present in the Supplementary text (section S2, "Analysis of the ALMA data") and also reported in [12,13], artificially produced features which mimic true atmospheric lines can be produced when analyzing data with such characteristics.

Many of these large fluctuations can be introduced by the particular parameters used in calibration during data reduction. After the publication of G2020 we notified the ALMA NAASC team of substantial differences in the final spectra when applying different treatments of the bandpass calibrations. Specifically, the type of bandpass calibration solution (traditional channel-to-channel vs. polynomial-fitted vs. smoothed), and in particular, enabling the "usescratch=True" setting in CASA's `setJy` for the model used for Callisto (the bandpass calibrator). The newly reprocessed data from the Joint ALMA Office (JAO) take these issues into considerations and are used by G2021a/b. We analyzed the data as presented originally in G2020 and could not recover a $PH_3$ signature (see Figure 1B), and we also reduced the ALMA data using the new JAO scripts as the starting point. We employed three separate procedures by three different groups (NASA/GSFC, Berkeley, and NRAO), making use of different methods in order to evaluate the significance of the reported ALMA $PH_3$ signatures. The three methods are: 1) employing the updated JAO scripts throughout (using CASA only, done at NASA/GSFC); 2) employing the updated JAO scripts, but proceeding independently beyond self-calibration (using CASA only; done at Berkeley) – note that phase-only self-calibration was done but not amplitude self-calibration, because the latter can be problematic for extended sources[9]; 3) as (2), but with the post-JAO-script reduction done in AIPS (using both CASA and AIPS, done at NRAO). We investigated including/removing certain baselines, with further details about the methods can be found in the methods section and in Supplementary text, section S2. All these analyses used the common foundation of the revised JAO scripts to do the initial calibration (in particular, the bandpass and complex gain vs. time calibrations), and very similar steps in the further data reduction (phase-only self-calibration, continuum subtraction, forming the image cube, and extracting the disk-averaged spectrum).

The quality of the bandpass calibration was improved with the updated JAO scripts, yet the residual spectrum still showed notable large fluctuations (we found that a 6$^{th}$ order polynomial captured most of the residual large fluctuations for method 1; G2020 considered a 12$^{th}$ order). Our other two re-analyses (2 and 3) led to practically flat spectra only needing 2$^{nd}$ order polynomial baselines (more information on the origin of these differences is explained in the Supplementary Text, section S2). Ultimately, all our analyses of the data using these different approaches and methodologies reveal no conclusive signature of $PH_3$ (Figure 1), leading to an upper-limit of < 1 ppbv (3$\sigma$) when employing a linewidth of 0.186 cm$^{-1}$/atm. When employing other proposed linewidths, the upper-limit would be $PH_3$ < 0.7 ppbv (linewidth of 0.12 cm$^{-1}$/atm) or < 1.5 ppbv (linewidth of 0.286 cm$^{-1}$/atm). This upper-limit is also consistent with other recently reported upper-limits derived from IR ground-based observations ($PH_3$ < 5 ppbv)[14] and from spacecraft data ($PH_3$ < 0.2 ppbv)[15]. We further validated our analysis of the ALMA data by analyzing other $SO_2$ and HDO lines in the same dataset (see Supplementary text, section S3).

**JMCT analysis and $SO_2$ contamination**: The analysis of JCMT spectral data in G2020 involved fitting of multiple polynomials for the purpose of continuum subtraction: an initial 4$^{th}$ order polynomial was removed, then a 9$^{th}$ order polynomial was removed from a smoothed spectrum, and finally a masked 8$^{th}$ order polynomial was removed. Recently, Thompson[13] explored the robustness of the two detection methods used in G2020, namely low order polynomial fits and higher order multiple polynomial fits, and found that neither line detection method is able to recover a statistically significant detection at the position of the $PH_3$/$SO_2$ line.

We investigate whether a weak $PH_3$ signature, if present, could be distinguished from $SO_2$ contamination, and what its contribution to the detected signal could be. To explore this, we employed the same VIRA45 temperature/pressure (T/P) profile used by G2020 (their Extended Data Figure 8) and the G2020 $SO_2$ profile presented in their Extended Data Figure 9 (~100 ppbv in the 70-90 km region, dropping to <30 ppbv above 95 km). If this potential $SO_2$ contaminant signature is removed from the JCMT data, then the original feature would be confined within the noise (Fig. 2). We next modelled a mesospheric $SO_2$ profile as measured by Venus Express using solar occultation[16], and similar to case D of ref [17] ($SO_2$ abundance of ~30 ppbv at 80 km, increasing to ~100 ppb at 90 km and reaching ~300 ppbv at 95 km). In this case, the $SO_2$ contamination signature is even stronger and fully captures the claimed $PH_3$ residual in G2020 and G2021a/b (Fig. 2). A comprehensive compilation of $SO_2$ measurements[12-13], including thousands of measurements from the Venus Express orbiter in UV and IR wavelengths as well as ground-based observations[16,18–27], show that $SO_2$ abundances at 80 km are frequently well in excess of 100 ppbv, with peak abundances occasionally exceeding 1000 ppbv. This shows that the $SO_2$ profiles we adopted are well within usual ranges of variability observed on Venus (more information about the considered profiles can be found in the Supplementary Material Section S1, "Vertical Profiles"). Having shown that the absorption line can be reproduced by reasonable vertical profile of $SO_2$, we therefore argue that this absorption line cannot be definitively attributed to $PH_3$.

**Probing altitude:** We also explored the altitudes from which these absorptions originate, which can be used to constrain photochemical models and facilitate comparison to other measurements[28,29]. This is ultimately related to the spectroscopic parameters of the targeted line and that of the competing radiative active species in this spectral region. When considering the linewidth of 0.186 cm$^{-1}$/atm for $PH_3$, at 70 km (3.4x10$^{-2}$ atm) the line would be 213 km/s wide, i.e., much broader than the narrow window region of ±5 km/s or ±10 km/s searched for $PH_3$. As a polynomial is fitted and subtracted, removing any spectral line information beyond this core region, the spectral information at broader widths (originating from lower altitudes) is thus removed. Our models only predict an observable $PH_3$ absorption at this frequency only when phosphine is present above 75 km (see details in the Supplementary Material Section S4, "Altitude of the probed narrow molecular absorptions"), and therefore these data provide no constraints on its abundance in the cloud deck (50-70 km), in contrast to G2020 and G2021a/b.

This can be seen in Fig. 2, a flat spectrum is calculated for the $PH_3$ profile determined from the chemical modelling described in G2020, in which $PH_3$ is present only below 70 km altitude.

**Conclusions**: By our independent analysis of the ALMA data, we set a limit of 1 ppbv for $PH_3$. We also show that the observed JCMT feature could be attributed to mesospheric $SO_2$ gas. Furthermore, for any $PH_3$ signature to be recoverable in either ALMA or JCMT analyses, $PH_3$ needs to be present at altitudes above 75 km. We conclude that the recent identification of $PH_3$ in the upper atmosphere of Venus is not supported.

**Acknowledgements and data availability**

This paper makes use of the 2018.A.00023.S ALMA data, available at https://almascience.nrao.edu/asax. We would like to commend the G2020 team for making their data and scripts available. ALMA is a partnership of ESO (representing its member states), NSF (USA) and NINS (Japan), together with NRC (Canada), MOST and ASIAA (Taiwan), and KASI (Republic of Korea), in cooperation with the Republic of Chile. The Joint ALMA Observatory is operated by ESO, AUI/NRAO and NAOJ. The JCMT data was collected under project S16BP007, and available at https://www.eaobservatory.org/jcmt/science/archive. The JCMT is operated by the EAO on behalf of NAOJ; ASIAA; KASI; CAMS as well as the National Key R&D Program of China (No. 2017YFA0402700). Additional funding support is provided by the STFC and participating universities in the UK and Canada. The authors wish to recognize and acknowledge the very significant cultural role and reverence that the summit of Maunakea has always had within the indigenous Hawaiian community. We are most fortunate to have the opportunity to conduct observations from this mountain.

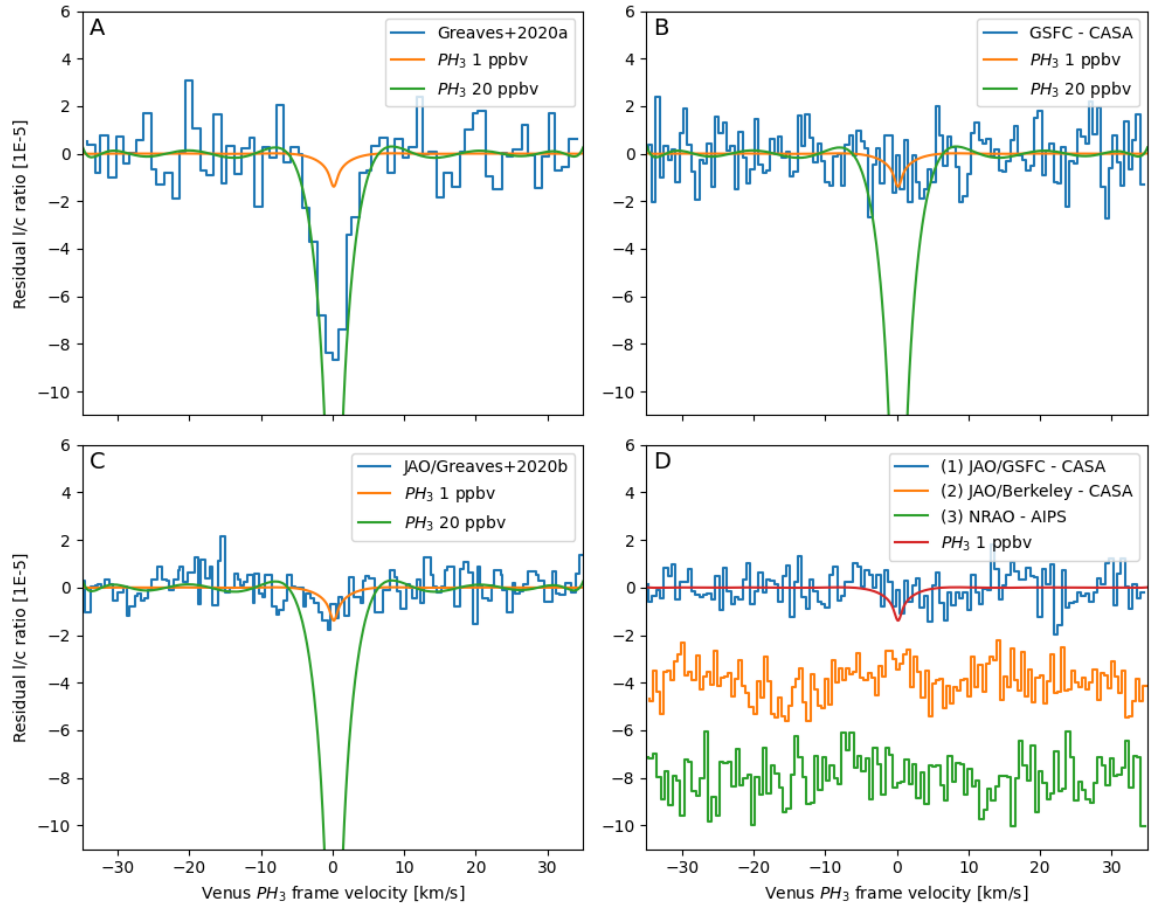

**Figure 1:** Comparison between the ALMA data as presented in G2020 and G2021a and our independent analyses of the same data retrieved from the ALMA science archive. Models with constant 1 and 20 ppbv PH$_3$ abundances are superposed. **Panel A**: The ALMA data as presented in Figure 2 of G2020 for the whole planet (dv:1 km/s). **Panel B:** Our original analysis of the raw ALMA data (dv:0.55 km/s), employing the original G2020 scripts but enabling the "usescratch=True" setting in CASA's `setJy`, correcting the bandpass calibration. **Panel C**: The re-analyzed data presented in G2021a using updated scripts, yet G2020 employed a higher (12$^{th}$ order) polynomial in the bandpass calibration (instead of 3$^{rd}$ order in the JAO script), and excluded short baselines (<33 m) from their analysis (dv: 0.55 km/s). **Panel D**: Our independent analysis of the JAO-revised ALMA data, employing different methods, a resolution of dv:0.55 km/s, methods 1 and 2 including all baselines and method 3 excluding <33 m baselines. A 6$^{th}$ order polynomial was removed for (1), see details in the methods section and in the supplementary section S2, while only a 2$^{nd}$ was removed for (2) and (3). The red line corresponds to the model with a constant abundance of 1 ppbv PH$_3$.

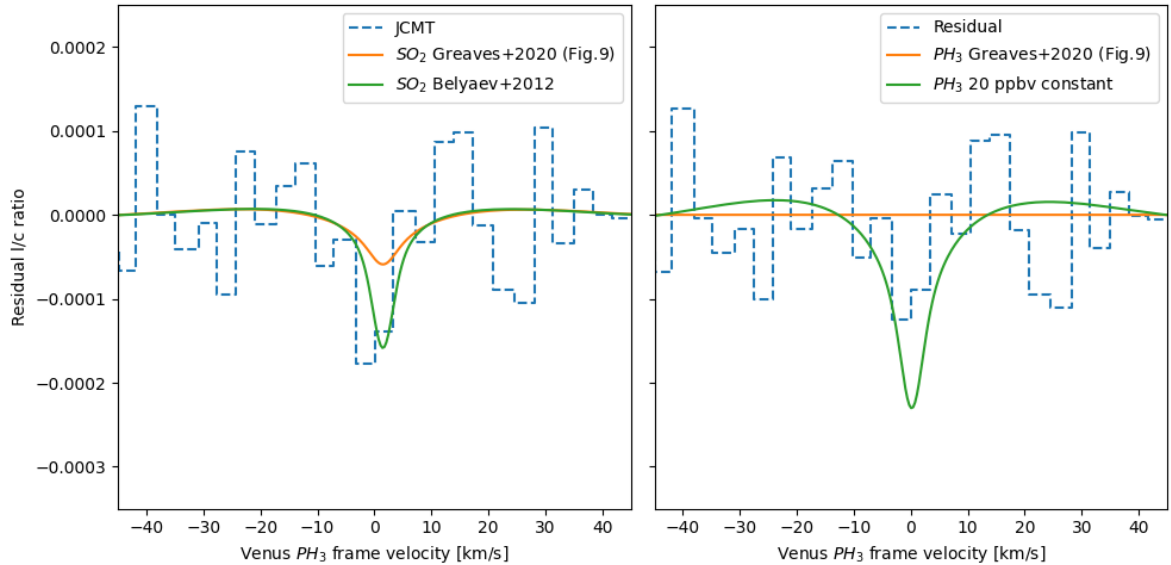

**Figure 2:** Comparison between the residual JCMT data as presented in Figure 1 of G2020 and models of $SO_2$ and $PH_3$. **Left**: The JCMT data for their mid-range solution with masking within ± 5 km/s, while "$SO_2$ Greaves+2020" is a model spectrum synthesized using the T/P in G2020/Fig. 8 and the $SO_2$ profile in G2020/Fig. 9. "$SO_2$ Belyaev+2012" is a typical profile in Belyaev+2012[16] (Fig. 9) and similar to Lincowski+2021[17] (case D), with an $SO_2$ abundance of ~30 ppbv at 80 km, and increasing to ~100 ppbv at 90 km and reaching ~300 ppb at 95 km. **Right**: "Residual" is obtained by removing the $SO_2$ signature from the JCMT data as modelled with a 3 km/s resolution using the G2020/Fig. 9 $SO_2$ profile. The "$PH_3$ Greaves+2020" spectrum is a model spectrum employing the $PH_3$ profile in G2020/Fig. 9 (peaking at 60 km with 20 ppbv) and the default linewidth (0.186 cm$^{-1}$/atm), while a model considering a vertically constant mixing ratio of 20 ppbv is labelled "$PH_3$ 20 ppbv constant". The "$PH_3$ Greaves+2020" is flat and featureless because these measurements only sample $PH_3$ above 75 km (see S4).

## Methods

We have employed three independent radiative transfer models: the Planetary Spectrum Generator (PSG, https://psg.gsfc.nasa.gov)[4], the Non-linear optimal Estimator for MultivariatE spectral analySIS (NEMESIS)[5] and the CfA planetary modeling tool[6]. The PSG radiative transfer analysis included the latest HITRAN $SO_2$ line parameters for a $CO_2$ atmosphere[7], a layer-by-layer, line-by-line study, and a full disk sampling scheme with 10 concentric rings. The NEMESIS analysis was also performed in line-by-line mode, and used the same spectroscopic data with a 5-point Gauss-Lobatto disc-integration scheme. As described in G2020, there is some uncertainty in the line-shape parameters for the $PH_3$ line in a $CO_2$ atmosphere. HITRAN reports an air linewidth of 0.067 cm$^{-1}$/atm, which would correspond to 0.12 cm$^{-1}$/atm in a $CO_2$ atmosphere if the value is scaled by the typical 1.8 scaling ratio observed for the $SO_2$ lines[8]. G2020 considered a theoretical estimate of 0.186 cm$^{-1}$/atm, and an upper range of 0.286 cm$^{-1}$/atm measured for the $NH_3$(J=1-0) line in $CO_2$ at 572.498160 GHz. Considering the uncertainty on this parameter, we adopt the same value in G2020 by default (0.186 cm$^{-1}$/atm).

When reducing the ALMA data, we downloaded the raw data and scripts as provided in the archive. By three separate reduction methods, we then used the provided scripts along with well-established techniques in the calibration and imaging of radio interferometric data to obtain final disk-averaged spectra for Venus. There are three scripts provided with the data: 1 – a calibration script; 2 – an imaging preparation script; 3 – an imaging script. The calibration script calculates and applies several system calibrations and flags data known to be invalid, then sets the model for Callisto and uses that for flux density scale and bandpass calibration (using a 3$^{rd}$ order polynomial fit to the bandpass amplitude), then does a complex gain vs. time calibration using the calibrator J2000-1748. The imaging preparation script Doppler shifts the spectral axis of the various spectral windows, then images Venus and uses that image to do phase-only self-calibration, then sets the model for Venus and does amplitude self-calibration.

The imaging script then makes a continuum image, does continuum subtraction, and makes the image cube. In both the continuum image and the image cube a primary beam correction is made, since the primary beam at these frequencies (with FWHM ~24") resolves the disk of Venus (diameter ~15.2" at the time). We note that the amplitude self-calibration step in the imaging preparation script can lead to problems when observing bright objects with sharp boundaries (like Venus or Jupiter)[9], and it also leads to the flux density scale being set incorrectly when using these scripts (because it will set the model flux density to that expected for Venus with no primary beam attenuation, but the primary beam actually attenuates that flux density by about 15%, and that attenuation is actually corrected for in the imaging script). All three of our independent analyses used the entire calibration script as provided from the archive. Our first analysis continued with the scripts as provided from the archive throughout, attempting to recreate as closely as possible the analysis in G2020 and G2021a/b. Our second analysis took the data after the doppler shift step, then proceeded with the reduction within CASA. The only notable difference from the imaging preparation and imaging scripts is the use of a limb-darkened model of Venus for the phase-only self-calibration, and not doing amplitude self-calibration at all. Our third analysis was nearly identical to the second, taking the data after the doppler shift step, but then proceeding within AIPS instead of CASA. The differences in this third analysis were addition of a second bandpass calibration using a model of Callisto, use of the same limb-darkened model of Venus as in the second analysis for the phase-only self-calibration, no use of amplitude self-calibration, and a scaling of the flux densities by the ratio of the expected (primary beam-attenuated) flux density from Venus to that calculated with a fit to the visibilities.  Aside from these differences, other steps in our second and third analyses were essentially the same as what was in the provided scripts. For the third analysis, baselines with separations less than 33 m were excluded when making the final image cube; for the first analysis, this was investigated and found to reduce the amplitude of the bandpass calibration artifacts, while increasing the (channel-to-channel) spectral noise. No $PH_3$ detection was evident in either case. For all three analyses, we used the entire disk of Venus to create the final spectrum.

**Supplementary material**

**S1: Vertical profiles**

The temperature profile as presented in "extended Data Figure 8" of G2020a is consistent with the Venus International Reference Atmosphere (VIRA)[30] for mid-latitudes (45 degrees latitude), and also consistent with later observations[31], so is also used in our simulations. The $SO_2$ profile presented in "extended Data Figure 9" of G2020a is consistent with previous observations[21–23], in particular at altitudes above 70 km relevant to this investigation, and employed in Figure 1 to model a potential $SO_2$ contamination signature. $SO_2$ is known to vary significantly in the mesosphere, with peak abundances beyond 1000 ppbv and a minimum at ~80 km in the 10-100 ppbv range[16,18–27]. Orbital missions, which provide global monitoring, observe large variations on timescales of hours to months superimposed on a long-term trend. Mesospheric $SO_2$ abundances at the time of the JCMT observations in June 2017 thus cannot be constrained by $SO_2$ abundances measured using ALMA in March 2019, and beam dilution could hinder the detectability of $SO_2$ in Venus with ALMA[17,32]. We also explored other plausible and reported mesospheric $SO_2$ profiles as measured by spacecraft[16] and similar to case D of ref [17], with $SO_2$ of ~30 ppbv at 80 km, and increasing to ~100 ppb at 90 km and reaching ~300 ppb at 95 km. See synthetic spectra for this case in Figure 2.

**S2: ALMA bandpass corrections**

From our first analysis, we note that the quality of the bandpass did improve with the updated JAO scripts, yet the residual spectra still show large fluctuations (see Supplementary Figure 1). In order to correct for this, a $12^{th}$ order polynomial for the amplitude bandpass calibration was employed in G2020b. We, however, preserved the calibration scheme in the JAO scripts ($3^{rd}$ order polynomial for amplitude bandpass calibration), and removed a $6^{th}$ order polynomial baseline fit from the final spectrum. The impact of masking the center line region and fitting a high-degree polynomial baseline can be quite problematic, as revealed by the effective disappearance of most of the $PH_3$ signature between G2020a and G2020b and demonstrated in supplementary figure 2. This was also shown systematically and independently for both the JCMT[13] data and the ALMA[12] data, which both revealed no $PH_3$ signature. We explored different polynomial fits to the residual data, testing the sensitivity of masking the line center region. We only observe an absorption feature as reported in G2020b when employing a $12^{th}$ order polynomial, yet a $6^{th}$ order already appears to provide a relatively good match to the fluctuations in the spectra (Supplementary Figure 1). We finally note that our second and third analyses result in notably flatter final spectra, only needing a $2^{nd}$ order polynomial to be fit and removed.

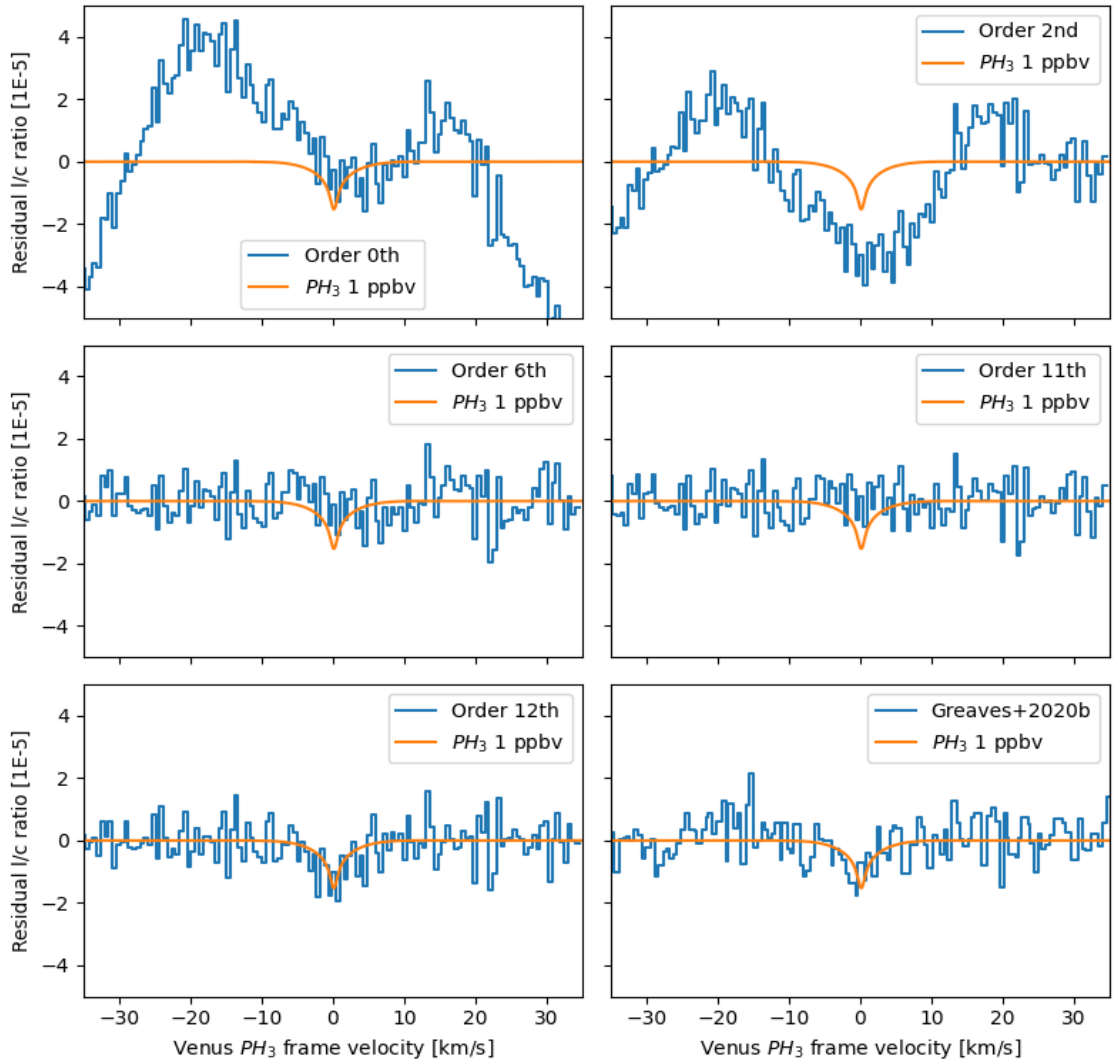

**Supplementary Figure 1:** By masking the center of the spectrum and by fitting a high-degree order polynomial around the line center, features in the core region may be artificially produced (see Supplementary Figure 2). In this figure, we present residual spectra as derived using the JAO scripts, and as presented in G2020b (see bottom/right panel). G2020b employed a bandpass polynomial of 12$^{th}$ order for the amplitude bandpass calibration, instead of the original 3$^{rd}$ order in the JAO scripts. We preserved all parameters as determined by the JAO team, included all baselines, and ultimately obtained the residual as presented in the top/left panel (order 0$^{th}$ baseline removed). We then explored sequentially higher degrees (while masking the center ± 5 km/s) and reached a reasonably good residual at order 6$^{th}$. The residual with a polynomial of order 11$^{th}$ is quite similar to the 6$^{th}$ order case, and only when employing a 12$^{th}$ order polynomial do we observe a feature as reported in G2020b. In the case of G2020b, the 12$^{th}$ order polynomial was not removed post-processing as shown here, but applied to the visibility bandpass calibration.

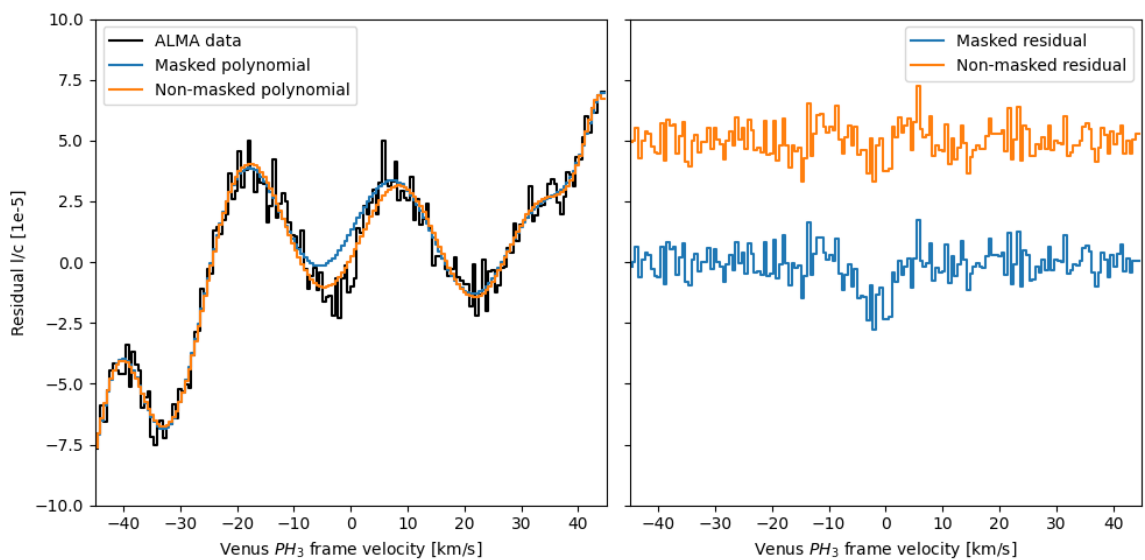

**Supplementary Figure 2:** Impact of masking the core region while fitting a high-degree polynomial order around the center velocities. **Left:** Continuum subtracted residual spectrum of Venus derived employing the original G2020a scripts (yet disabling self-calibration[9] and including all baselines). Polynomial fits (12$^{th}$ order) are superimposed: "non-masked" trace is a fit to the entire spectrum, while "masked" is a fit excluding the center ± 5 km/s. **Right**: Spectra after subtracting the polynomial fits - The non-masked residual has been offset for clarity. These residuals are derived employing modified scripts as those employed by G2020a, and they are solely to show the complexities on fitting a high-degree polynomial to ripply data.

### S3: Validation of the ALMA analysis by interpreting other nearby lines

We independently analyzed the ALMA data using our calibration scripts for the region near the $SO_2$ line at 267.537458 GHz and the HDO (J=$2_{2,0}$-$3_{1,3}$) line at 266.16107 GHz (see Fig. Supplementary Figure 3). We do not detect $SO_2$, and estimate a mesospheric abundance <10 ppbv, which is consistent with the reported range of variability observed of mesospheric $SO_2$ (see S1). Furthermore, beam dilution could hinder the detectability of $SO_2$ in Venus[17,32]. We have a detection of HDO, and when assuming a D/H of 200[1] we estimate a mesospheric value of ~60 ppbv for water, in agreement with previous findings[23].

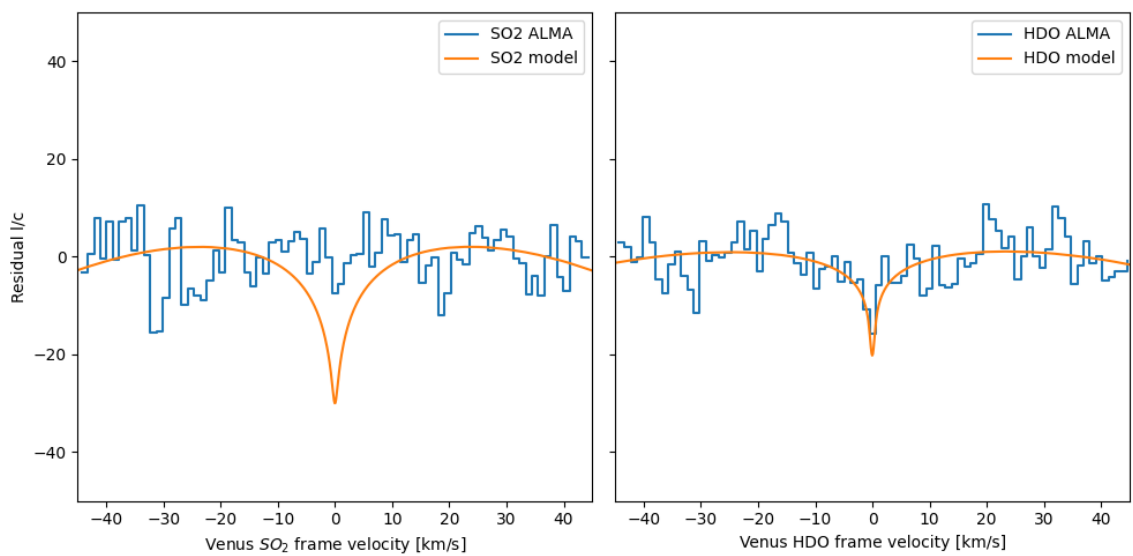

**Supplementary Figure 3:** Comparison between models and ALMA data, for the $SO_2$ ($J=13_{3,11}-13_{2,12}$) transition at 267.537458 GHz and for the HDO ($J=2_{2,0}-3_{1,3}$) transition at 266.161070 GHz. **Left**: Our independently processed ALMA data for the $SO_2$ line, and "$SO_2$ model" is a synthetic spectrum modeled employing the VIRA45 T/P profile and mesospheric (70-90 km) abundance of $SO_2$ of 10 ppb. **Right**: Our analysis of the ALMA data for a nearby HDO line, and "HDO model" is a synthetic spectrum as modeled adopting a D/H of 200[20], and a plausible $H_2O$ abundances of ~60 ppbv in the mesosphere (70-100 km). A 2$^{nd}$ order polynomial was removed from each spectrum (while masking the ± 5 km/s region).

### S4: Altitude of the probed narrow molecular absorptions

The altitude from which a specific absorption/emission originates is related to the spectroscopic parameters of the targeted line and those of competing radiatively-active species in this spectral region. Going deeper into the Venusian atmosphere, the pressure increases, and so does the Lorentzian width of the lines. For instance, when considering the linewidth of 0.186 cm$^{-1}$/atm for $PH_3$, at 70 km ($3.4\times10^{-2}$ atm) the line would be 213 km/s at 267 GHz, much broader than the narrow window region of ± 5 km/s or ± 10 km/s used to search for $PH_3$. For the spectral range shown in Figures 1/2 and in the figures in G2020a/b/c, the spectral line would appear completely flat, since it encompasses a spectral range much broader than shown. Since a polynomial is removed for pixels beyond the core region, any $PH_3$ information beyond this velocity would also be removed. A $PH_3$ linewidth of 20 km/s would occur at $3.2\times10^{-3}$ atm (81 km), so in principle information below this altitude would be removed. Furthermore, as we go deeper into the atmosphere, collision-induced-absorptions and the broad wings of other strong submillimeter absorbers (e.g., $CO_2$, $SO_2$, $H_2O$) dominate over the $PH_3$ signatures, further masking any potential signatures in these regions.

In order to quantify the specific altitude in which $PH_3$ would produce a detectable absorption in the residual data, we synthetized spectra using the linewidth considered in G2020a/b (0.186 cm$^{-1}$/atm) at different altitudes for a 10 km layer of $PH_3$. The model spectra for each altitude are shown in Supplementary Figure 4, with the bottom altitude listed in the y-axis. The figure demonstrates that these observations only sample $PH_3$, if present, above 75 km in altitude. The other linewidth suggested in G2020a/b of 0.286 cm$^{-1}$/atm requires an absorption region even higher in the atmosphere (>80 km). This is consistent with the findings and modeling presented in Lincowski+2021[17].

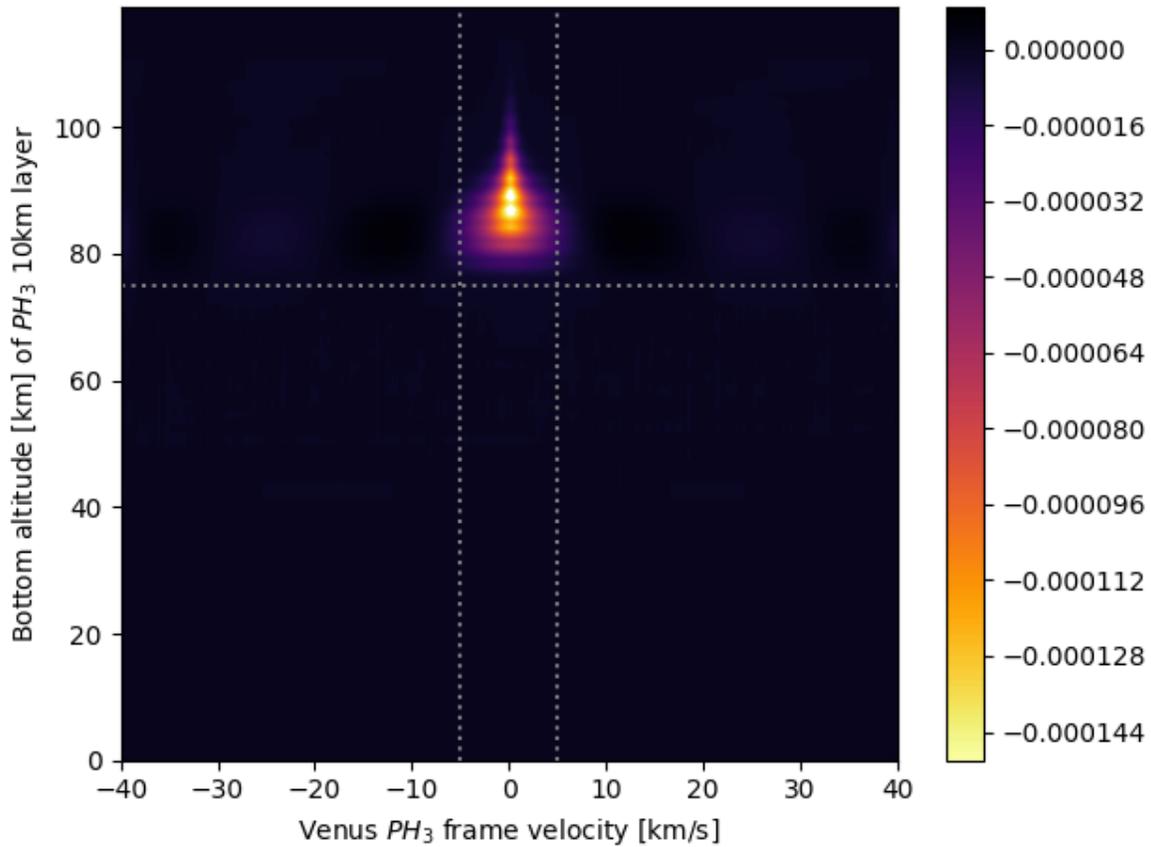

**Supplementary Figure 4:** Sensitivity analysis of the recoverable $PH_3$ J=1-0 signature as a function of velocity and altitude on Venus. This was computed by synthesizing spectra of a hypothetical 10 km-deep layer of $PH_3$ located at different altitudes. The vertical axis defines the lower bound of the layer, with the horizontal dotted line marking the 75 km altitude. The vertical lines denote the ± 5 km/s window around the line center of the transition. A 6$^{th}$ order polynomial was removed for each spectrum, while also masking the ± 5 km/s center region, as performed to the ALMA data presented in Figure 2, Supplementary Figures 1 and 2, and the figures in G2020a/b. A linewidth of 0.186 cm$^{-1}$/atm was considered as in G2020a/b.